\documentclass[a4paper,english,final,twocolumn,showpacs,pra,lengthcheck]{revtex4}
\usepackage{ae,aecompl}
\usepackage[T1]{fontenc}
\usepackage[latin9]{inputenc}
\usepackage{textcomp}
\usepackage{amsmath}
\usepackage{graphicx}
\usepackage{amssymb}

\makeatletter

\pdfpageheight\paperheight
\pdfpagewidth\paperwidth

\@ifundefined{textcolor}{}
{%
 \definecolor{BLACK}{gray}{0}
 \definecolor{WHITE}{gray}{1}
 \definecolor{RED}{rgb}{1,0,0}
 \definecolor{GREEN}{rgb}{0,1,0}
 \definecolor{BLUE}{rgb}{0,0,1}
 \definecolor{CYAN}{cmyk}{1,0,0,0}
 \definecolor{MAGENTA}{cmyk}{0,1,0,0}
 \definecolor{YELLOW}{cmyk}{0,0,1,0}
 }

\usepackage{textcomp}
\usepackage{babel}

\usepackage{amsfonts}

\providecommand{\U}[1]{\protect\rule{.1in}{.1in}}
\makeatletter
\providecommand{\LyX}{L\kern-.1667em\lower.25em\hbox{Y}\kern-.125emX\@}
\@ifundefined{textcolor}{}
{
\definecolor{BLACK}{gray}{0}
\definecolor{WHITE}{gray}{1}
\definecolor{RED}{rgb}{1,0,0}
\definecolor{GREEN}{rgb}{0,1,0}
\definecolor{BLUE}{rgb}{0,0,1}
\definecolor{CYAN}{cmyk}{1,0,0,0}
\definecolor{MAGENTA}{cmyk}{0,1,0,0}
\definecolor{YELLOW}{cmyk}{0,0,1,0}
}
\makeatother
\makeatother

\makeatother

\usepackage{babel}

\makeatother

\makeatother

\usepackage{babel}

\begin{document}

\preprint{This line only printed with preprint option}

\title{Localization effects induced by decoherence in superpositions of
many-spin quantum states}

\author{Gonzalo A. \'Alvarez}

\altaffiliation{galvarez@e3.physik.uni-dortmund.de}

\affiliation{Fakult\"at Physik, Technische Universit\"at Dortmund, Dortmund, Germany.}

\author{Dieter Suter}

\altaffiliation{Dieter.Suter@tu-dortmund.de}

\affiliation{Fakult\"at Physik, Technische Universit\"at Dortmund, Dortmund, Germany.}


\begin{abstract}
The spurious interaction of quantum systems with their environment
known as decoherence leads, as a function of time, to a decay of coherence
of superposition states. Since the interactions between system and
environment are local, they can also cause a loss of spatial coherence:
correlations between spatially distant parts of the system are lost
and the equilibrium states can become localized. This effect limits
the distance over which quantum information can be transmitted, e.g.,
along a spin chain. We investigate this issue in a nuclear magnetic
resonance quantum simulator, where it is possible to monitor the spreading
of quantum information in a three-dimensional network: states that
are initially localized on individual spins (qubits) spread under
the influence of a suitable Hamiltonian apparently without limits.
If we add a perturbation to this Hamiltonian, the spreading stops
and the system reaches a limiting size, which becomes smaller as the
strength of the perturbation increases. This limiting size appears
to represent a dynamical equilibrium. We present a phenomenological
model to describe these results. 
\end{abstract}
\maketitle

\section{\textit{\emph{Introduction}}}

Controlling quantum mechanical systems has received increasing attention
in recent years \cite{ladd_quantum_2010}, mainly because we are starting
to be able to perform computations with quantum mechanical systems.
This has the potential of solving computational problems for which
no efficient algorithm exists for classical computers \cite{Shor1994,DiVincenzo1995,Nielsen00}.
However, we will only be able to realize this potential if we learn
to control large quantum systems with high reliability. The control
of \textit{small} quantum systems has been thoroughly explored in
the last years \cite{ladd_quantum_2010}, however the study and control
of \textit{large} quantum systems has not been tackled so far. The
reason for this is partly the difficulty of simulating large quantum
systems on classical computers, which limits the number of qubits
to about 20 when the system is in a pure state \cite{Raedt04,zhang_modelling_2007}.
Additionally, it is also possible to calculate the dynamics of mixed
states if the initial state is localized by using quantum parallelism
of a single pure state evolution \cite{popescu_entanglement_2006,alvarez_quantum_2008}.
On the other side, the present primitive state of quantum computation
only allows incomplete control of large quantum states. So far, the
only physical system that offers this possibility is nuclear magnetic
resonance (NMR) of dipolar coupled spins \cite{Suter04,Suter06,Lovric2007}.
In particular, the main problems are the lack of individual addressing
of qubits and decoherence. The latter degrades the quantum information
of a given state \cite{Zurek03}. Decoherence was shown to increase
as the size of the quantum system increases, making the largest systems
the most sensitive to perturbations \cite{Suter04,Suter06,Krojanski2006,Cory06,Lovric2007,sanchez_time_2007}.
While reducing the perturbation strength typically reduces decoherence,
in complex and large systems, the decoherence time can be independent
of the perturbation strength over a range of coupling strengths \cite{PhysicaA,JalPas01}.

Decoherence is well known as a process that causes the decay of quantum
information. Avoiding or reducing decoherence is thus the main ingredient
for implementing large scale quantum computers. Several techniques
have been proposed for this purpose, including dynamical decoupling
\cite{5916}, decoherence-free subspaces \cite{3045}, and quantum
error correction \cite{3921,6581}. These proposals have been tested
on small systems of nuclear spins \cite{6013}, trapped ions \cite{monz:200503}
or spin model quantum memories \cite{biercuk_optimized_2009,Du2009}.
Some of these techniques were successfully applied to large quantum
systems with thousands of qubits where their decoherence time was
extended by almost two orders of magnitude \cite{Suter06,sanchez_time_2007}.

Decoherence not only affects the survival time of quantum information,
it also affects the distance over which it can be transmitted \cite{Pomeransky2004,Chiara2005,Keating2007,Burrell2007,Apollaro2007,Allcock2009,alvarez_nmr_2010,alvarez_decoherence_2010,Zwick2011}.
For example, spin systems, and in particular spin chains, can be used
to transfer quantum information over large distances \cite{Bose2003,Christandl2004}.
This kind of systems were studied with liquid state NMR for small
numbers of spins \cite{Madi97,nielsen_complete_1998,zhang_simulation_2005,zhang_iterative_2007,Alvarez_perfect_2010}
and slightly larger numbers by solid-state NMR \cite{Cappellaro2007,Rufeil-Fiori2009,Zhang2009}.
However, once decoherence is considered, in the simplest two spin
quantum channel, it was shown that when the effective system-environment
interaction exceeds a given strength, it becomes impossible to perform
even a simple SWAP operation \cite{JCP06,alvarez_decoherence_2010}.
Instead of having the expected oscillatory transfer of a state going
forth and back between the two spins, an overdamped dynamics due to
the appearance of a localized state appears at a critical value or
exceptional point of the perturbation strength \cite{JCP06,SSC07,rotter_non-hermitian_2009,rotter_environmentally_2010,rotter_role_2010}.
Similarly, this could be observed in a 3-spin chain. If one spin is
suffering a perturbation, when it exceeds a given strength the dynamics
localizes in the remaining two spins \cite{alvarez_signatures_2007}.
In a more general situation with longer spin-chains, it was recently
pointed out that imperfections or disorder of the spin-coupling that
drives the state transfer can induce localization of the quantum information
\cite{Pomeransky2004,Chiara2005,Burrell2007,Keating2007,Apollaro2007,Allcock2009,Zwick2011}
in a process related to Anderson localization \cite{Anderson1958,anderson_local_1978}.
In more complex 3D spin-network topologies, we demonstrated experimentally
a similar behavior by studying the localization effects induced by
the finite precision of quantum gate operations used for transferring
quantum states \cite{alvarez_nmr_2010}.

In this paper, we extend our previous work \cite{alvarez_nmr_2010},
where we prepared the system first as individual, uncorrelated spins
and measured the build-up of clusters of correlated spins of increasing
size. Introducing a perturbation to the Hamiltonian that generates
these clusters, we find that the size of the clusters reaches an upper
bound. This upper bound appears to be a dynamic equilibrium: if the
cluster size is initially larger than this equilibrium value, it decreases
under the effect of the perturbed Hamiltonian, while the unperturbed
Hamiltonian leads to an increase. The equilibrium size decreases with
increasing strength of the perturbation. For these experiments, we
use a solid-state NMR quantum simulator. While this system does not
allow addressing of individual qubits, it represents an excellent
test-bed for studying different aspects of decoherence and information
transfer.

\section{\textit{\emph{The quantum simulator}} }

We consider a system of equivalent spins $I=1/2$ in the presence
of a strong magnetic field. The Hamiltonian of the system is \begin{align}
\widehat{\mathcal{H}} & =\widehat{\mathcal{H}}_{z}+\widehat{\mathcal{H}}_{dd}\nonumber \\
 & =\omega_{z}\sum_{i}\hat{I}_{z}^{i}+\sum_{i<j}d_{ij}\left[2\hat{I}_{z}^{i}\hat{I}_{z}^{j}-(\hat{I}_{x}^{i}\hat{I}_{x}^{j}+\hat{I}_{y}^{i}\hat{I}_{y}^{j})\right],\end{align}
 where $\omega_{z}$ is the Larmor frequency, $d_{ij}$ the coupling
constants and $\hat{I}_{x}^{i},\hat{I}_{y}^{i},\hat{I}_{z}^{i}$ are
the spin operators that can be represented by $\hat{I}_{u}^{i}=\frac{1}{2}\hat{\sigma}_{u}$
with $\hat{\sigma}_{u}$ Pauli operators. $\widehat{\mathcal{H}}_{z}$
represents the Zeeman interaction and $\widehat{\mathcal{H}}_{dd}$
the dipolar interaction \cite{Slichter}. The latter is truncated
to commute with the strong Zeeman interaction ($\omega_{z}\gg d_{ij}$),
assuming that the effects of its non-commuting part are negligible.
In a frame of reference rotating at the Larmor frequency \cite{Slichter},
the Hamiltonian of the spin system reduces to $\widehat{\mathcal{H}}_{dd}$.

The quantum simulations start from the high-temperature thermal equilibrium
\cite{Slichter}. Using the notation $\hat{I}_{z}=\sum_{i}\hat{I}_{z}^{i}$,
we can write the thermal equilibrium state as\begin{align}
\rho_{0} & \approx\left(\hat{1}+\frac{\hbar\omega_{z}}{k_{\mathrm{B}}T}\hat{I}_{z}\right)/\mathrm{Tr}\left\{ \hat{1}\right\} .\end{align}
 The unity operator $\hat{1}$ commutes with all operators, including
the Hamiltonian and the density operator. It does not contribute to
the observable signal and it is therefore convenient to exclude it
from the density operator. The resulting density operator for the
initial state of the system is then $\hat{\rho}_{0}\propto\hat{I}_{z}$.
In this state, the spins are uncorrelated and the density operator
commutes with the Hamiltonian $\widehat{\mathcal{H}}_{dd}$.

We performed all experiments on a home-built solid state NMR spectrometer
with a $^{\text{1}}$H resonance frequency of $\omega_{z}/2\pi=300$
MHz. The spins are the protons of polycrystalline adamantane, where
the average strength of the average dipolar interaction, determined
from the width of the resonance line, is $7.9$ kHz.

\section{Growth of spin clusters}

\subsection{Cluster size}

The spin clusters that we consider can be written in product operator
form as \[
\hat{I}_{u}^{l}...\hat{I}_{v}^{o}\hat{I}_{w}^{p}\left(u,v,w=x,y,z\right).\]
 Here, the indexes $l,o,p$ identify the spins involved in the given
cluster. We write $K$ for the number of terms in this product, \emph{i.e.},
for the number of spins in the cluster. Experimentally, we generate
these clusters using an NMR method developed by Pines and coworkers
\cite{5105,Baum1985}. It is based on generating an average Hamiltonian
$\widehat{\mathcal{H}}_{0}$ that does not commute with the thermal
equilibrium state

\begin{align}
\widehat{\mathcal{H}}_{0} & =-\sum_{i<j}d_{ij}\left[\hat{I}_{x}^{i}\hat{I}_{x}^{j}-\hat{I}_{y}^{i}\hat{I}_{y}^{j}\right]\nonumber \\
 & =-\frac{1}{2}\sum_{i<j}d_{ij}\left[\hat{I}_{+}^{i}\hat{I}_{+}^{j}+\hat{I}_{-}^{i}\hat{I}_{-}^{j}\right],\label{flip-flip}\end{align}
where $\hat{I}_{\pm}^{j}=\left(\hat{I}_{x}^{j}\pm i\hat{I}_{y}^{j}\right)$.
In the usual computational basis, we write the states as $\left|M,n_{M}\right\rangle $
where $M$ is the total magnetic quantum number, \emph{i.e.}, $\hat{I}_{z}\left|M,n_{M}\right\rangle =M\left|M,n_{M}\right\rangle $,
and $n_{M}$ distinguishes different states with the same $M$. The
Hamiltonian (\ref{flip-flip}) flips simultaneously two spins with
the same orientation. Starting from the thermal equilibrium state,
it generates a density operator where only elements $\rho_{ij}$ with
\[
\Delta M=M(i)-M(j)=2n,\, n=0,1,2\dots\]
 are populated. The index $i$ and $j$ refer to the different spin
states in the Zeeman basis as described above. Such a density operator
element $\rho_{ij}$ is called a $\Delta M$ quantum coherence. Off-diagonal
elements with $\Delta M=0$ represent zero-quantum coherences and
diagonal elements correspond to populations.

\begin{figure}
\includegraphics[bb=50bp 70bp 680bp 570bp,clip,width=1\columnwidth]{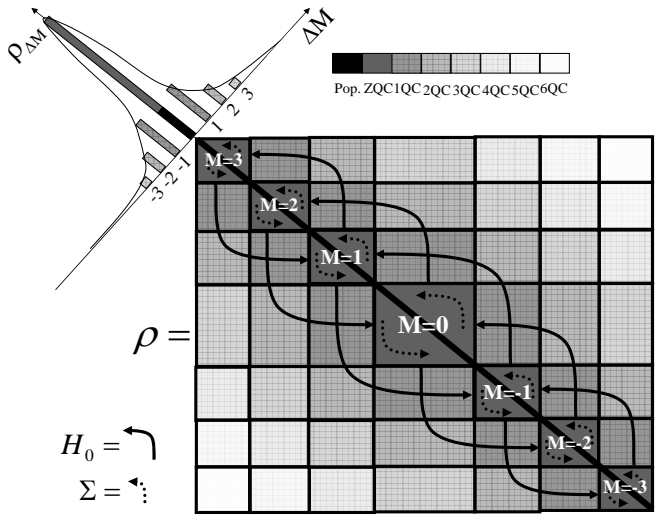}

\caption{Density matrix in the computational basis. The black diagonal line
represents the populations (Pop. in the legend). The diagonal blocks
(dark gray) represent coherences of $\hat{\rho}$ with $\Delta M=0$
(ZQC) and the non-diagonal blocks are the $\Delta M$ multiple-quantum
coherence (MQC) blocks (gray tones). The solid and dashed arrows indicate
the effect of $\mathcal{\widehat{H}}_{0}$ and the perturbation interaction
$\Sigma=\mathcal{\widehat{H}}_{dd}$ respectively on the thermal equilibrium
density operator. A schematic representation of a typical distribution
is given by the Gaussian-like shape on the top-left corner.}

\label{Flo:DM_scheme_evolution} %
\end{figure}

Figure \ref{Flo:DM_scheme_evolution} shows a graphical representation
of the density matrix in the computational basis. The black diagonal
line represents the population elements of the density matrix. The
diagonal blocks (gray) represent coherences of $\hat{\rho}$ with
$\text{\ensuremath{\Delta}}M=0$ and moving away from these diagonal
blocks, the $\Delta M$ multiple-quantum coherence (MQC) order increases
and they are represented by different gray values. The solid arrows
indicate the effect of $\mathcal{\widehat{H}}_{0}$ on the thermal
equilibrium density operator. Adding the elements of $\hat{\rho}$
with a given $\Delta M$, we obtain a distribution function of the
MQC elements of the density matrix $\hat{\rho}_{\Delta M}$. A schematic
representation of its distribution is given by the Gaussian-like shape
on the upper-left corner of Fig. \ref{Flo:DM_scheme_evolution}. This
distribution is initially a delta function on the diagonal of $\hat{\rho}$
and spreads with time.

To determine the average number of correlated spins, the technique
relies on the fact that in a system of $K$ spins, the number of transitions
with a given $\Delta M$ follows a binomial distribution \cite{Wokaun1978,Hoffman1970}
\begin{equation}
n\left(\Delta M,K\right)=\frac{\left(2K\right)!}{\left(K+\Delta M\right)!\left(K\text{\textminus}\Delta M\right)!}.\end{equation}
 For $K\gg1,$ the binomial distribution can be well approximated
with a Gaussian\begin{equation}
n\left(\Delta M,K\right)\propto\exp\left(\text{\textminus}\frac{\Delta M^{2}}{K}\right),\label{eq:gaussian}\end{equation}
 whose half-width at $e^{-1}$ is $\sigma=\sqrt{K}$ . Thus, we can
determine the effective size of the spin clusters in a given state
by measuring the distribution of the MQCs of its density operator
$\rho$ as a function of the coherence order $\Delta M$.

\subsection{Growth}

\begin{figure}
\includegraphics[bb=0bp 0bp 612bp 792bp,clip,scale=0.3]{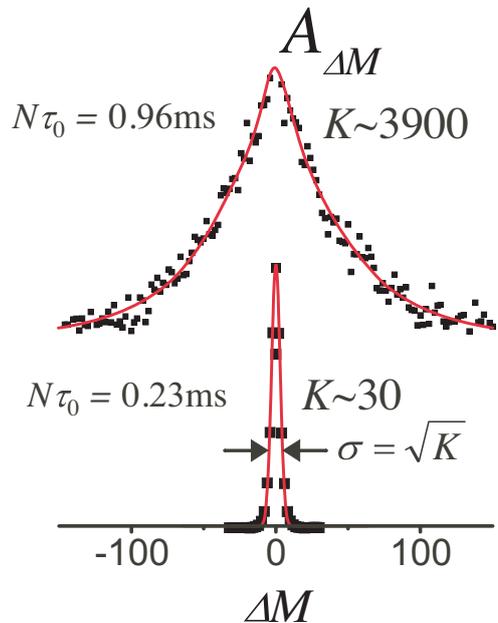}

\caption{(Color online) Multiple-quantum coherence (MQC) distributions for
two different evolution times. The cluster-size is extracted from
the variance $\sigma=\sqrt{K}$. }

\label{Flo:AMEvolution} %
\end{figure}

Figure \ref{Flo:AMEvolution} shows two distribution functions of
the MQCs for two different durations $N\tau_{0}$ of the evolution
under $\mathcal{\widehat{H}}_{0}$. They clearly demonstrate the increasing
width of the MQC distribution and thereby the increasing size of the
spin clusters. %

\begin{figure}
\includegraphics[bb=100bp 590bp 420bp 765bp,clip,width=7.5cm]{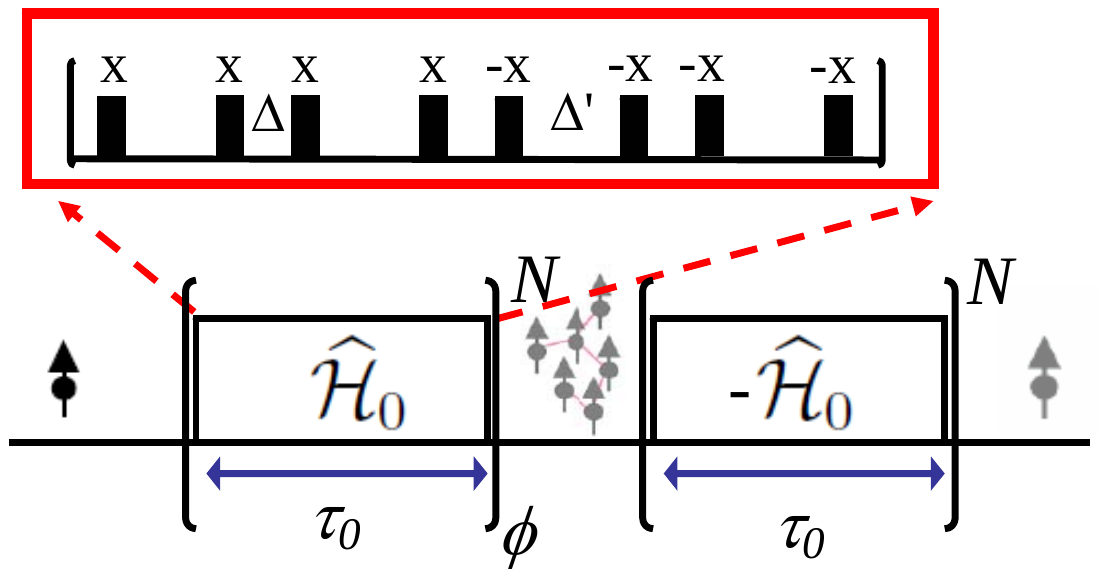}

\caption{(Color online) NMR sequence for generating large spin clusters. The
effective Hamiltonian $\widehat{\mathcal{H}}_{0}$ is generated by
the sequence of $\text{\ensuremath{\pi}/2}$ pulses shown in the upper
part of the figure, where $\Delta^{\prime}=2\Delta+\tau_{p}$ and
$\tau_{p}$ is the $\text{\ensuremath{\pi}/2}$ pulse duration \cite{Baum1985}.}

\label{Flo:NMRseqH0} %
\end{figure}

The elements of the density operator with a given $\Delta M$ can
be distinguished by rotating the system around the $z-$axis. A rotation
$\hat{\phi}_{z}=e^{-i\phi\hat{I}_{z}}$ by $\phi$ changes the density
operator as

\begin{equation}
\hat{\rho}\left(\phi,t\right)=\hat{\phi}_{z}\hat{\rho}\left(t\right)\hat{\phi}_{z}^{\dagger}=\sum_{\Delta M}\hat{\rho}_{\Delta M}^{{}}\left(t\right)e^{i\Delta M\phi},\label{eq:rhophi}\end{equation}
 where $\hat{\rho}\left(t\right)=e^{-i\widehat{\mathcal{H}}_{0}t}\hat{\rho}_{0}e^{i\mathcal{\widehat{H}}_{0}t}$
and $\hat{\rho}_{\Delta M}(t)$ contains the elements that involve
coherences of order $\Delta M$. The experimental observables in NMR
are $\hat{I}_{x}$, $\hat{I}_{y}$, and by means of a $\pi/2$ pulse,
also $\hat{I}_{z}$. They are operators with $\Delta M=\pm1$ or $\Delta M=0$,
respectively. Thus we are not able to measure all the elements of
the density matrix. In order to quantify the $\hat{\rho}_{\Delta M}^{{}}$
blocks, we use the pulse sequence shown in Fig. \ref{Flo:NMRseqH0}.%

The system initially evolves for a period of duration $N\tau_{0}$
under the Hamiltonian $\left(\mathcal{\widehat{H}}_{0}\right)_{\phi}=\hat{\phi}_{z}\mathcal{\widehat{H}}_{0}\hat{\phi}_{z}^{\dagger}$,
\emph{i.e.}, \begin{align}
\hat{\rho}_{0}\xrightarrow{\left(\mathcal{\widehat{H}}_{0}\right)_{\phi}N\tau_{0}}\hat{\rho}_{\phi}\left(N\tau_{0}\right) & =\hat{\phi}_{z}\hat{\rho}\left(N\tau_{0}\right)\hat{\phi}_{z}^{\dagger}\nonumber \\
 & =\hat{\phi}_{z}e^{-i\widehat{\mathcal{H}}_{0}N\tau_{0}}\hat{\rho}_{0}e^{i\mathcal{\widehat{H}}_{0}N\tau_{0}}\hat{\phi}_{z}^{\dagger}\nonumber \\
 & =\sum_{\Delta M}\hat{\phi}_{z}\hat{\rho}_{\Delta M}^{{}}\left(N\tau_{0}\right)\hat{\phi}_{z}^{\dagger}\nonumber \\
 & =\sum_{\Delta M}\hat{\rho}_{\Delta M}^{{}}\left(N\tau_{0}\right)e^{i\Delta M\phi}.\end{align}
 A subsequent evolution period with the same duration $N\tau_{0}$
under $-\mathcal{\widehat{H}}_{0}$, which causes an evolution backward
in time, yields the final density operator\begin{multline}
\xrightarrow{\left(-\mathcal{\widehat{H}}_{0}\right)N\tau_{0}}\hat{\rho}_{f}\left(2N\tau_{0}\right)=e^{i\widehat{\mathcal{H}}_{0}N\tau_{0}}\hat{\rho}_{\phi}\left(N\tau_{0}\right)e^{-i\widehat{\mathcal{H}}_{0}N\tau_{0}}\\
=\sum_{\Delta M}\left[e^{i\mathcal{\widehat{H}}_{0}N\tau_{0}}\hat{\rho}_{\Delta M}^{{}}\left(N\tau_{0}\right)e^{-i\mathcal{\widehat{H}}_{0}N\tau_{0}}\right]e^{i\Delta M\phi}.\end{multline}
 If $\hat{I}_{z}$ is the NMR observable, we obtain the signal \begin{align}
S\left(\phi,N\tau_{\mbox{0}}\right) & =\mbox{Tr}\left\{ \hat{I}_{z}\hat{\rho}_{f}\left(2N\tau_{\mbox{0}}\right)\right\} \nonumber \\
 & =\mathrm{Tr}\left\{ e^{-i\mathcal{\widehat{H}}_{0}N\tau_{0}}\hat{\rho}_{0}e^{i\mathcal{\widehat{H}}_{0}N\tau_{0}}\hat{\rho}_{\phi}\left(N\tau_{0}\right)\right\} \nonumber \\
 & =\mathrm{Tr}\left\{ \hat{\rho}\left(N\tau_{0}\right)\hat{\rho}_{\phi}\left(N\tau_{0}\right)\right\} \nonumber \\
 & =\sum_{\Delta M}\mbox{\ensuremath{e^{i\phi\Delta M}}Tr}\left\{ \hat{\rho}_{\Delta M}^{2}\left(N\tau_{\mbox{0}}\right)\right\} \label{eq:unperturbed_signal}\\
 & =\sum_{\Delta M}\mbox{\ensuremath{e^{i\phi\Delta M}}}A\left(\Delta M\right),\end{align}
 where $A(\Delta M)$ are the amplitudes of the MQ spectrum. To extract
these amplitudes from the experimental data, we measure the signal
$S\left(\phi,N\tau_{\mbox{0}}\right)$ as a function of $\phi$ and
perform a Fourier transform with respect to $\phi$. We quantify the
cluster size $K$ by inverting the relation $A\left(\Delta M\right)=n\left(\Delta M,K\right)$
(see Ref. \cite{Baum1985} for details).

\begin{figure}
\includegraphics[bb=70bp 50bp 710bp 560bp,clip,width=1\columnwidth]{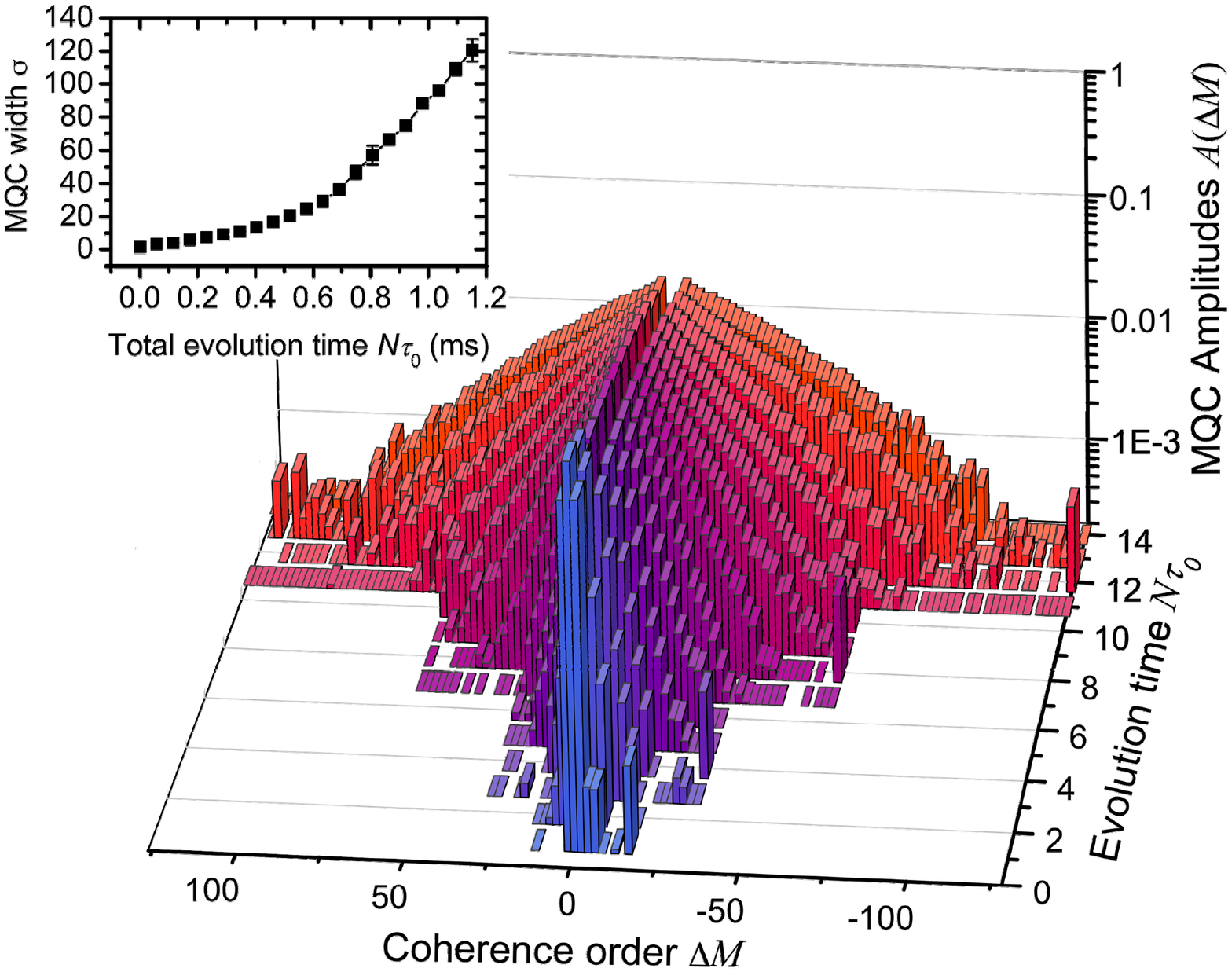}

\caption{(Color online) Time evolution of the multiple quantum coherence distribution.
The main panel shows the time evolution of the MQC spectrum $A(\Delta M)$.
The inset shows the time evolution of $\sigma$, where $A(\sigma,N\tau_{0})/A(0,N\tau_{0})=1/e$. }

\label{Flo:Free_evolution} %
\end{figure}

\begin{figure}
\includegraphics[bb=0bp 360bp 612bp 775bp,clip,width=1\columnwidth]{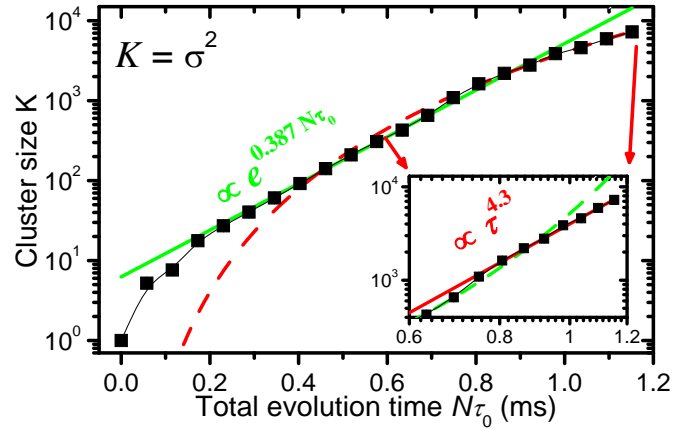}

\caption{(Color online) Time evolution of the cluster size of correlated spins
with the Hamiltonian $\widehat{\mathcal{H}}_{0}$ (black square points).
The main panel shows it in log scale manifesting the region where
the growth is exponential and the inset shows it in log-log scale
manifesting its power law growth for long times.}

\label{fig:Growth} %
\end{figure}

If the system evolves under the Hamiltonian (\ref{flip-flip}), the
width of the $A\left(\Delta M\right)$ distribution increases indefinitely,
as shown in Fig. \ref{Flo:Free_evolution}. The main panel shows the
MQC distributions $A\left(\Delta M\right)$ for different evolution
times. We quantify the spreading of the MQC distribution by measuring
their widths ($\sigma$) at $A(\sigma,N\tau_{0})/A(0,N\tau_{0})=1/e$
for different evolution times. The inset of the figure shows the time
evolution of $\sigma$. Equivalently, its corresponding cluster size
of correlated spins also grows indefinitely, as shown in Figure \ref{fig:Growth}.%

The Hamiltonian of Eq. (\ref{flip-flip}) is prepared by means of
the standard NMR sequence \cite{5105,Baum1985} shown in the upper
part of Fig. \ref{Flo:NMRseqH0}. This sequence consists of $\pi$/2
rotations of the spins separated by periods of free precession, $\Delta$
and $\Delta^{\prime}$, under the dipolar Hamiltonian $\widehat{\mathcal{H}}_{dd}$
and it was shown to approximate quite well the ideal Hamiltonian $\widehat{\mathcal{H}}_{0}$.
It is inverted experimentally by shifting the phase of all RF pulses
by $\pm\pi/2$ \cite{5105}. We used pulse durations $\tau_{\pi/2}=2.8\mu\mathrm{s}$
and delays $\Delta=2\mu\mathrm{s}$ and $\Delta^{\prime}=2\Delta+\tau_{p}$
giving a cycle time $\tau_{0}=57.6\mu\mbox{s}$. %

\subsection{MQC distributions and cluster-size}

In this section we summarize previous works that described the MQC
distribution, its evolution and how it is related with the cluster-size
of correlated spins in order to compare their results with our experimental
observation. So far, it has not been possible to derive a consistent
theoretical model of the processes involved in this type of many-body
system. The most accepted and simplest model for interpreting the
MQC distributions was proposed by Baum \emph{et al.} \cite{Baum1985}.
It assumes equal excitation probabilities (for a given system size)
for every coherence term of the density matrix and therefore predicts
 a Gaussian distribution whose variance is related to the number of
correlated spins. This is the usually adopted model because it provides
the simplest qualitative description of the MQC evolution. However,
the  experimentally observed distributions are not always Gaussian
\cite{Lacelle1993}. This is also the case in our experiments: the
data shown in Fig. \ref{Flo:Free_evolution} are better described
by an exponential distribution $A(\Delta M)$. Our results are therefore
more consistent with the distribution discussed by Lacelle \emph{et
al.} \cite{Lacelle1993}, who predicted an exponential MQC distribution
and found agreement with earlier experimental results from adamantane.
Additionally, they predicted a dynamical scale invariance in the growth
process of the MQC\textcompwordmark{}.

Due to the difficulty of treating many-spin systems rigorously, quantum
mechanical predictions for the MQC profile shapes and their interpretations
are still missing. For example, if the MQC profile is not a Gaussian
distribution, it is not trivial how to extract from it the cluster-size.
While, most of previous works studying the distribution of MQCs have
assumed Gaussian distributions, it is more convenient to use a parameter
that is independent of the MQC distribution form. Khitrin proposed
to use the second moment of the MQC spectrum for this purpose \cite{Khitrin1997}.
This parameter was shown to be related to the second moment of the
system Hamiltonian, in our case given by Eq. (\ref{flip-flip}) and
$K(t)$ is proportional to the second moment of the MQC distribution.

To determine the cluster-size from our experimental data we measured
the half-width $\sigma=\sqrt{K}$ of the MQC distributions at $1/e$.
We chose $\sigma=\sqrt{K}$ because an exponential MQC distribution
of the form $\exp\left(-\Delta M/\sqrt{K(t)}\right)$ has the same
width as the Gaussian distribution (\ref{eq:gaussian}). Assuming
this distribution, its second moment is $2K$ while for a Gaussian
distribution it is $K/2$. With the exception of the Gaussian model
for a MQC distribution, there is no rigorous model that provides the
exact factor that converts the second moment of the distribution to
the cluster size $K$. However, they are of the same order of magnitude
and thus the error of the cluster-size determination is a scale factor
of order 1. In our experiments, the MQC distribution has the same
shape during the complete range of the experimentally accessible evolution
time. Accordingly, any error of the determination of the cluster-size
is thus a constant factor independent of the value of $K$ and this
does not change the conclusions of this article.

The growth depends on the spin-coupling network topology as was observed
experimentally \cite{Baum1985,Baum1986,Lacelle1991,Lacelle1993,Khitrin1997,zobov_second_2006,Zobov2008}.
Mainly three types of growth were observed: (i) indefinite growth,
(ii) localized growth and (iii) localized growth at a first stage
(intra-molecular-like localization) and then a further indefinite
growth (inter-molecular).

Results for adamantane fall into the first category. Some models suggest
a power law for the growth \cite{Lacelle1991,Gleason1993} and these
preliminary results \cite{Lacelle1991,Lacelle1993} seem to match
with an effective 3D spin-coupling network topology. That 3D behavior
seems to be achieved only for cluster sizes above 1000 spins \cite{Lacelle1991}.
However, more recent works suggest an exponential growth of the cluster
size when many neighboring spins are contained in the spin-coupling
topology \cite{zobov_second_2006,Zobov2008}. This is the case of
adamantane and the experiments agree well with that prediction.

Experimentally, the data shown in Fig. \ref{fig:Growth} suggest three
stages of the cluster size evolution: (i) An initial period where
the evolution cannot be described as exponential or power law, (ii)
a period of exponential growth, and (iii) a power law regime. During
the first stage, the MQC distribution changes from a Gaussian-like
distribution to an exponential distribution. Thus, the determination
of the cluster size could be affected by two kinds of systematic errors:
(i) at small cluster sizes, the small number of correlated spins may
make the statistical assumptions questionable and (ii) the change
of distribution from Gaussian to exponential may change the appropriate
scale factors. The exponential growth agrees with the behavior expected
by theoretical predictions for this kind of system \cite{zobov_second_2006,Zobov2008}.
And finally the power law behavior could be due to the following two
reasons: (i) as predicted by Lacelle \cite{Lacelle1991} after a certain
number of spins (around 1000 in adamantane) the effective spin-network
topology turns into a 3D spin-coupling network, which leads to a power
law growth or (ii) because the experimental generation of $\widehat{\mathcal{H}}_{0}$
contains non-idealities, they can be considered as a perturbation
that generates localization effects, as was demonstrated in Ref. \cite{alvarez_nmr_2010}
and discussed in the following sections. Something similar to the
latter point could be interpreted from the predictions of Ref. \cite{Zobov2008}.

\section{Effect of Perturbation}

\subsection{Perturbed evolution}

The evolution under the Hamiltonian $\mathcal{\widehat{H}}_{0}$ can
be reversed completely by changing the Hamiltonian from $\widehat{\mathcal{H}}_{0}$
to $-\widehat{\mathcal{H}}_{0}$. If this time reversal is perfect,
the signal (\ref{eq:unperturbed_signal}) for phase $\phi=0$ is independent
of the evolution time $N\tau_{0}$, $\sum_{\Delta M}A(\Delta M,N\tau_{0})=\mathrm{const}$.

Experimentally, the Hamiltonian $\widehat{\mathcal{H}}_{0}$ is generated
as an effective Hamiltonian. Because of experimental imperfections,
it always deviates from the ideal Hamiltonian (\ref{flip-flip}).
As a result, the actual dynamics deviates from the ideal one and,
in particular, we cannot exactly invert the perturbed Hamiltonian
and thus revert the time evolution. Because of the deviations in $\widehat{\mathcal{H}}_{0}$,
the quantity $\sum_{\Delta M}A(\Delta M)$ is no longer conserved,
but decays with increasing evolution time. This decay is not uniform,
but it affects mostly those components of the density operator that
correspond to strongly delocalized coherence. As a result, the spreading
of the information is attenuated and the system becomes effectively
localized. It is this latter effect that we want to study here; we
isolate it from the overall decrease of the signal by normalizing
the MQ spectra such that the total signal $\sum_{\Delta M}A(\Delta M)$
for $\phi=0$ is again constant in time.

To analyze this deviation from the ideal evolution, we introduce a
perturbation $\widehat{\Sigma}$, whose strength we can control experimentally
and study the behavior of the system as a function of the perturbation
strength. We choose the dipole-dipole coupling for this perturbation,
$\widehat{\Sigma}=\mathcal{\widehat{H}}_{dd}$, which is a local interaction:
every spin interacts mostly with its nearest neighbors, while the
coupling strength with more distant spins drops off as $1/d^{3}$.

\begin{figure}
\includegraphics[bb=96bp 604bp 489bp 700bp,clip,width=7.5cm]{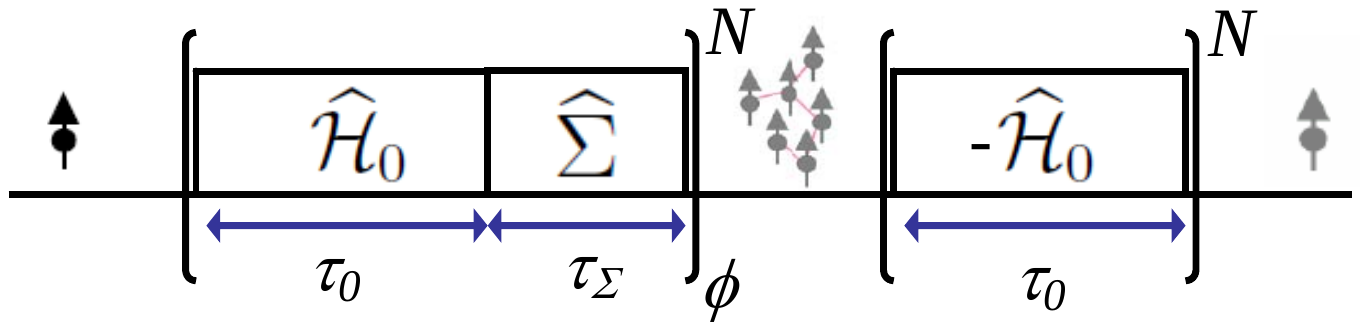}

\caption{(Color online) NMR sequence for the quantum simulations. A perturbed
evolution is achieved when $\tau_{\Sigma}\ne0$. The effective Hamiltonian
$\widehat{\mathcal{H}}_{0}$ is generated by the sequence of $\text{\ensuremath{\pi}/2}$
pulses shown in the upper part of Fig. \ref{Flo:NMRseqH0} and $\widehat{\Sigma}=\widehat{\mathcal{H}}_{dd}$
is the free evolution Hamiltonian. }

\label{fig:Periods} %
\end{figure}

We add this Hamiltonian to the ideal Hamiltonian $\widehat{\mathcal{H}}_{0}$
by concatenating short evolution periods under $\widehat{\mathcal{H}}_{dd}$
with evolution periods under $\widehat{\mathcal{H}}_{0}$. We label
the durations of the two time periods $\tau_{\Sigma}$ and $\tau_{0}$,
as shown in Fig. \ref{fig:Periods}.%

When the duration $\tau_{\mathrm{c}}=\tau_{0}+\tau_{\Sigma}$ of each
cycle is short compared to the inverse of the dipolar couplings $d_{ij}$,
the resulting evolution can be described by the effective Hamiltonian\begin{equation}
\widehat{\mathcal{H}}_{\mathrm{eff}}=(1-p)\widehat{\mathcal{H}}_{0}+p\widehat{\Sigma},\label{Heff}\end{equation}
 where the relative strength $p=\tau_{\Sigma}/\tau_{\mathrm{c}}$
of the perturbation $\widehat{\Sigma}=\widehat{\mathcal{H}}_{dd}$
can be controlled by adjusting the duration $\tau_{\Sigma}$. In the
experiment, we compare the artificially perturbed evolution of $\widehat{\mathcal{H}}_{\mathrm{eff}}$
with the $\widehat{\mathcal{H}}_{0}$ evolution with its intrinsic
errors. While the intrinsic errors reduce the signal or the overall
fidelity, they do not cause localization on the time scale of our
experiments (see Fig. \ref{fig:Growth}).

Considering now this perturbation, starting from thermal equilibrium,
the state of the system at the end of $N$ cycles is \begin{equation}
\hat{\rho}^{\mathcal{H}_{\mathrm{eff}}}\left(N\tau_{\mathrm{c}}\right)=e^{-i\widehat{\mathcal{H}}_{\mathrm{eff}}N\tau_{\mathrm{c}}}\hat{\rho}_{0}e^{i\widehat{\mathcal{H}}_{\mathrm{eff}}N\tau_{\mathrm{c}}}.\end{equation}
 Taking into account now the complete sequence of evolutions given
by Fig. \ref{fig:Periods}\begin{eqnarray}
\hat{\rho}_{0}\xrightarrow{\left(\mathcal{\widehat{H}}_{\mathrm{eff}}\right)_{\phi}N\tau_{\mathrm{c}}}\hat{\rho}_{\phi}^{\mathcal{H}_{\mathrm{eff}}}\left(N\tau_{\mathrm{c}}\right) & = & \hat{\phi}_{z}\hat{\rho}^{\mathcal{H}_{\mathrm{eff}}}\left(N\tau_{\mathrm{c}}\right)\hat{\phi}_{z}^{\dagger}\nonumber \\
 & = & \hat{\phi}_{z}e^{-i\widehat{\mathcal{H}}_{\mathrm{eff}}N\tau_{\mathrm{c}}}\hat{\rho}_{\mathrm{0}}e^{i\mathcal{\widehat{H}}_{\mathrm{eff}}N\tau_{\mathrm{c}}}\hat{\phi}_{z}^{\dagger}\nonumber \\
 & = & \sum_{\Delta M}\hat{\phi}_{z}\hat{\rho}_{\Delta M}^{\mathcal{H}_{\mathrm{eff}}}\left(N\tau_{\mathrm{c}}\right)\hat{\phi}_{z}^{\dagger}\nonumber \\
 & = & \sum_{\Delta M}\hat{\rho}_{\Delta M}^{{\mathcal{H}_{\mathrm{eff}}}}\left(N\tau_{\mathrm{c}}\right)e^{i\Delta M\phi}\end{eqnarray}
 \begin{multline}
\xrightarrow{\left(-\mathcal{\widehat{H}}_{0}\right)N\tau_{0}}\hat{\rho}_{f}^{\mathcal{H}_{\mathrm{eff}}}\left(N\tau_{\mathrm{c}}+N\tau_{0}\right)=\\
=e^{i\widehat{\mathcal{H}}_{0}N\tau_{0}}\hat{\rho}_{\phi}^{\mathcal{H}_{\mathrm{eff}}}\left(N\tau_{\mathrm{c}}\right)e^{-i\widehat{\mathcal{H}}_{0}N\tau_{0}}\\
=\sum_{\Delta M}\left[e^{i\mathcal{\widehat{H}}_{0}N\tau_{0}}\hat{\rho}_{\Delta M}^{\mathcal{H}_{\mathrm{eff}}}\left(N\tau_{\mathrm{c}}\right)e^{-i\mathcal{\widehat{H}}_{0}N\tau_{0}}\right]e^{i\Delta M\phi}.\end{multline}
 Thus the NMR signal, which is measured after the last backward evolution
$\exp\left\{ i\widehat{\mathcal{H}}_{0}N\tau_{0}\right\} $, can be
written as\begin{align}
S\left(\phi,N\tau_{0}+N\tau_{\mathrm{c}}\right) & =\mbox{Tr}\left\{ \hat{I}_{z}\hat{\rho}_{f}^{\mathcal{H}_{\mathrm{eff}}}\left(N\tau_{\mathrm{c}}+N\tau_{0}\right)\right\} \nonumber \\
 & =\mathrm{Tr}\left\{ e^{-i\mathcal{\widehat{H}}_{0}N\tau_{0}}\hat{\rho}_{0}e^{i\mathcal{\widehat{H}}_{0}N\tau_{0}}\hat{\rho}_{\phi}^{\mathcal{H}_{\mathrm{eff}}}\left(N\tau_{\mathrm{c}}\right)\right\} \nonumber \\
 & =\mathrm{Tr}\left\{ \hat{\rho}^{\mathcal{H}_{0}}\left(N\tau_{0}\right)\hat{\rho}_{\phi}^{\mathcal{H}_{\mathrm{eff}}}\left(N\tau_{\mathrm{c}}\right)\right\} \nonumber \\
 & =\sum_{\Delta M}\mbox{\ensuremath{e^{i\phi\Delta M}}Tr}\left\{ \hat{\rho}_{\Delta M}^{\mathcal{H}_{0}}\left(N\tau_{0}\right)\hat{\rho}_{\Delta M}^{\mathcal{H}_{\mathrm{eff}}}\left(N\tau_{\mathrm{c}}\right)\right\} \label{eq:M-fidelities}\\
 & =\sum_{\Delta M}\mbox{\ensuremath{e^{i\phi\Delta M}}}A(\Delta M).\end{align}
 Reinterpreting these expressions, we define the effective observable
$\mathcal{\widehat{A}}=e^{-i\mathcal{\widehat{H}}_{0}N\tau_{0}}\hat{\rho}_{0}e^{i\mathcal{\widehat{H}}_{0}N\tau_{0}}=\hat{\rho}^{\mathcal{H}_{0}}\left(N\tau_{0}\right)$,
where $\hat{\rho}^{\mathcal{H}_{0}}$ is the density operator of the
unperturbed evolution. The NMR signal becomes then $S(\phi,N\tau_{\mathrm{c}})=\mbox{Tr}\left\{ \mathcal{\widehat{A}}\hat{\rho}_{\phi}^{\mathcal{H}_{\mathrm{eff}}}\left(N\tau_{\mathrm{c}}\right)\right\} $.

\begin{figure}
\includegraphics[bb=185bp 60bp 740bp 530bp,clip,width=1\columnwidth]{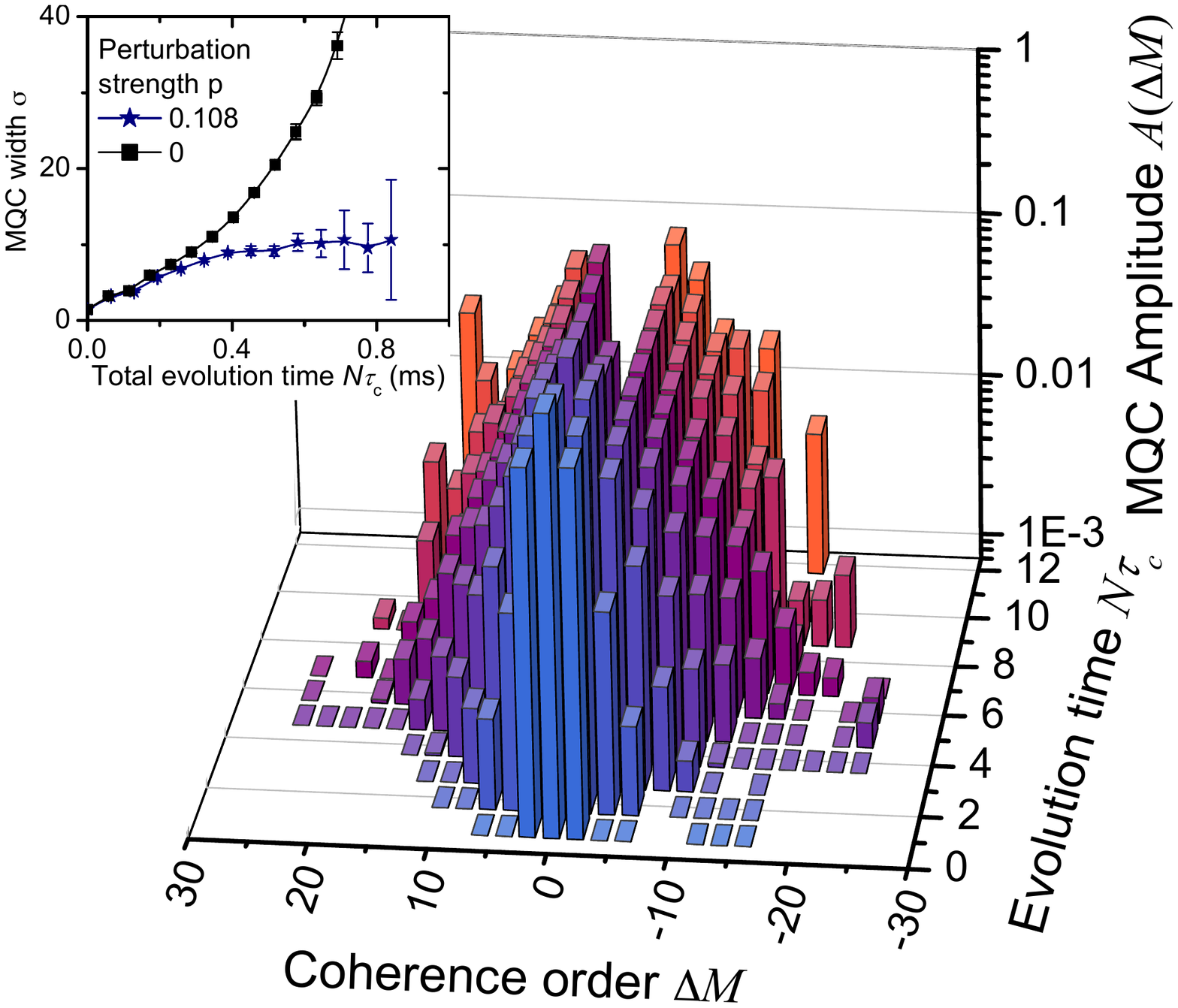}

\caption{(Color online) Time evolution of the multiple quantum coherence distribution
under a perturbation $p=0.108$. The main panel shows the time evolution
of the $M$QC spectrum $A(\Delta M)$. The inset shows the time evolution
of the standard deviation $\sigma$ for the unperturbed case compared
with the perturbed case. This comparison shows directly the localization.}

\label{Flo:MQCdistribution_perturbed} %
\end{figure}

For the ideal evolution ($p=0$), Eq. (\ref{eq:unperturbed_signal})
is recovered, where $A(\Delta M)$ correspond to the squared amplitudes
of the density operator elements $\hat{\rho}_{\Delta M}^{\mathcal{H}_{0}}\left(N\tau_{0}\right)$
with coherence order $\Delta M$. For the perturbed evolution, $(p\ne0)$,
they are reduced by the overlap of the actual density operator elements
$\hat{\rho}_{\Delta M}^{\mathcal{H}_{\mathrm{eff}}}\left(N\tau_{\mathrm{c}}\right)$
with the ideal ones. We extract these amplitudes by performing a Fourier
transformation with respect to $\phi$. Figure \ref{Flo:MQCdistribution_perturbed}
shows, as an example, the resulting $A(\Delta M)$ as a function of
time for $p=0.108$. The main panel represents the MQC distributions
for different evolution times. The main difference compared to Fig.
\ref{Flo:Free_evolution} is that the MQC spectrum does not spread
indefinitely \cite{alvarez_nmr_2010}. We consider this saturation
of the spreading of the MQC spectrum evolution as evidence of localization
due to the perturbation (see below). The localization effects are
easily visualized by directly comparing the generation of high-order
multiple quantum coherences of the unperturbed case of Fig. \ref{Flo:Free_evolution}
with the perturbed one of Fig. \ref{Flo:MQCdistribution_perturbed}.
While the distribution spreads continuously in Fig. \ref{Flo:Free_evolution},
it reaches a limiting value in Fig. \ref{Flo:MQCdistribution_perturbed}.
The inset of Fig. \ref{Flo:MQCdistribution_perturbed} shows the time
evolution of the width $\sigma$ for $p=0.108$ compared with the
unperturbed case.

\begin{figure}
\includegraphics[bb=0bp 270bp 570bp 730bp,clip,width=8cm]{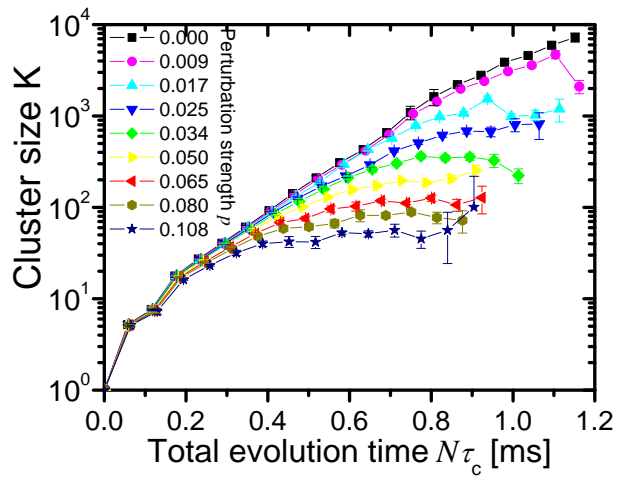}

\caption{(Color online) Time evolution of the cluster size for different perturbation
strengths. The black squares represent the unperturbed time evolution
and the other symbols correspond to the perturbed evolutions according
to the legend. }

\label{fig:Kvst} %
\end{figure}

From the width of the MQC spectrum, we calculate the size of the spin
clusters. The black squares of Fig. \ref{fig:Kvst} show the average
number of correlated spins as a function of time for an unperturbed
evolution, $p=0$. This is the log scale representation of the curve
of Fig. \ref{fig:Growth}. The other symbols of the figure show the
evolution of the number of correlated spins for different values of
$p$. Initially, the cluster size $K(N\tau_{\mathrm{c}})$ starts
to grow as in the unperturbed evolution, but then it saturates after
a time that decreases with increasing perturbation strength $p$.
As we explained above, we consider this as evidence of localization
induced by the perturbation and it is related to spatial localization.
The size of the cluster at which this saturation occurs is also determined
by the strength of the perturbation: increasing perturbation strength
reduces the limiting cluster size.

\subsection{Interpretation}

The terms in Eq. (\ref{eq:M-fidelities}) are related to the fidelity
of the corresponding density operator component $\hat{\rho}_{\Delta M}$
with respect to the corresponding component resulting from the unperturbed
evolution\begin{equation}
f_{\Delta M}=\frac{\mbox{Tr}\left\{ \hat{\rho}_{\Delta M}^{\mathcal{H}_{0}}\left(N\tau_{0}\right)\hat{\rho}_{\Delta M}^{\mathcal{H}_{\mathrm{eff}}}\left(N\tau_{\mbox{c}}\right)\right\} }{\mbox{Tr}\left\{ \hat{\rho}_{\Delta M}^{\mathcal{H}_{0}}\left(N\tau_{0}\right)\hat{\rho}_{\Delta M}^{\mathcal{H}_{0}}\left(N\tau_{0}\right)\right\} }\end{equation}
 which reaches unity for vanishing perturbation ($p=0$). We therefore
consider the reduction as a quantitative measure of the effect of
the perturbation. The decoherence of different $\Delta M$-blocks
of the density matrix was studied when a given quantum state prepared
by evolution under $\mathcal{\widehat{H}}_{0}$ evolves under a pure
$\mathcal{\widehat{H}}_{dd}=\widehat{\Sigma}$ perturbation \cite{Suter04,Krojanski2006}.
However, the quantification and characterization of how a perturbation
disturbs the different blocks of the density matrix during the creation
of spin-cluster states is still not known. This merits further studies
like in Ref. \cite{Doronin2011}. Here, we focus only on the effect
on the number of correlated spins and do not consider the change of
the overall amplitude. Thus we normalize the integral of each spectrum
and determine its width by fitting it to a Gaussian or an exponential.

The spreading of the MQC spectrum is generated by the effective Hamiltonian
created with the sequence of Fig. \ref{Flo:DM_scheme_evolution}.
The cluster size that we determine here corresponds to an overlap
of the actual state with the ideal state resulting from unperturbed
evolution. This is similar to the fidelity measure in quantum computing,
where the agreement between the actual state of the system with the
target state is measured. In our case, the target state is a growing
cluster, while the actual state grows only for some time until it
reaches a limiting size - typical for localization. The size of this
localized state decreases with increasing perturbation strength.

During our experiment, the magnetization is uniform throughout the
sample, so the process does not lead to a spatial redistribution of
magnetization. However, since we measure the number of correlated
spins, we can attribute a length scale to the resulting state. Given
a suitable initial state, the same process would generate highly entangled
multi-qubit states. In the presence of a perturbation, the number
of qubits that can be entangled in this way is limited to the size
of the resulting cluster.

\section{Evidence for a dynamical equilibrium cluster size}

\begin{figure}
\includegraphics[bb=0bp 570bp 605bp 770bp,clip,scale=0.4]{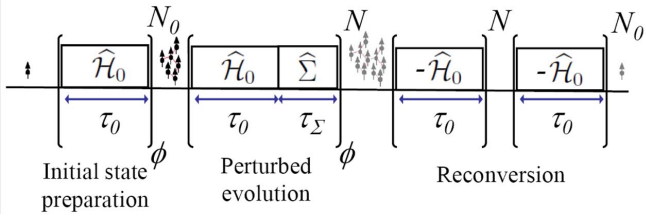}

\caption{(Color online) NMR pulse sequence for preparing different initial
clusters sizes and subsequently evolving them in the presence of a
perturbation. }

\label{Flo:NMRseqKini} %
\end{figure}

The experimental results presented above show that the cluster size
reaches a stationary value. It remains to be seen if this limiting
size results from a slow-down in the growth \cite{Burrell2007} or
it represents a dynamic equilibrium state. We showed that actually
they achieve a dynamical equilibrium state \cite{alvarez_nmr_2010}.
In order to do that we repeated the previous experiment for a series
of initial conditions corresponding to different clusters sizes. Figure
\ref{Flo:NMRseqKini} shows the corresponding pulse sequence: The
initial state preparation, consisting of an evolution of duration
$N_{0}\tau_{0}$ under the unperturbed Hamiltonian $\widehat{\mathcal{H}}_{0}$,
generates clusters of size $K_{0}$. During the subsequent perturbed
evolution of duration $N\tau_{\mathrm{c}}$, these initial clusters
evolve and Eq. (\ref{eq:M-fidelities}) becomes

\begin{multline}
S\left(\phi,N\tau_{\mathrm{c}}\right)=\sum_{\Delta M}\mbox{\ensuremath{e^{i\phi\Delta M}}}A(\Delta M)\\
=\sum_{\Delta M}\mbox{\ensuremath{e^{i\phi\Delta M}}Tr}\left\{ \hat{\rho}_{\Delta M}^{\mathcal{H}_{0}}\left(N\tau_{0},N_{0}\tau_{0}\right)\hat{\rho}_{\Delta M}^{\mathcal{H}_{\mathrm{eff}}}\left(N\tau_{\mbox{c}},N_{0}\tau_{0}\right)\right\} .\end{multline}

\begin{figure}
\includegraphics[bb=10bp 0bp 570bp 770bp,clip,scale=0.4]{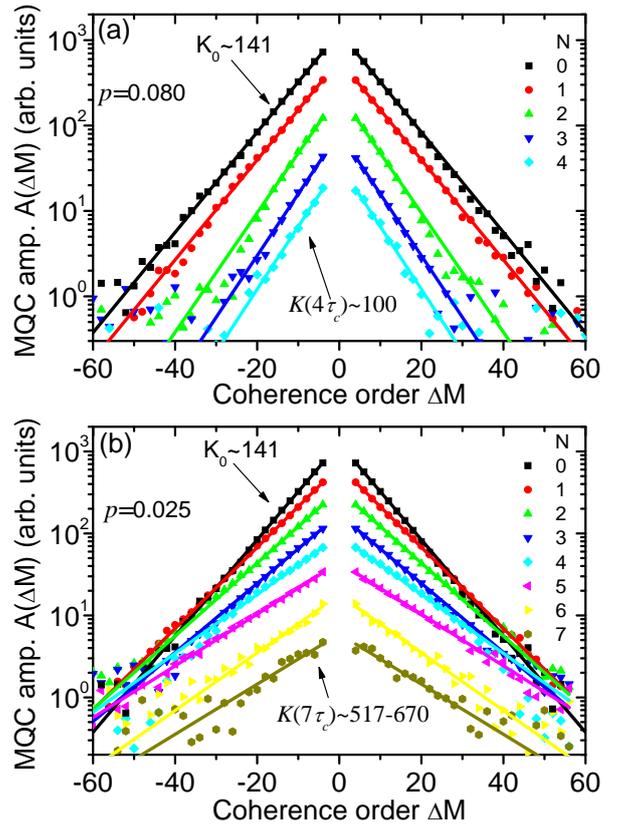}

\caption{(Color online) Time evolution of the multiple quantum coherence distribution
in the presence of a perturbation starting from an initial sate with
$K_{0}\simeq141$ correlated spins. The initial MQC distribution of
the initial state is given by the black squares and other colored
symbols gives its further time evolution for two different perturbation
strengths $p=0.080$ (Panel a) and $p=0.025$ (Panel b). }

\label{Flo:MQC_spectrums_dyn_eq} %
\end{figure}

Figure \ref{Flo:MQC_spectrums_dyn_eq} shows the results from these
measurements. The symbols represent the amplitudes $A(\Delta M)$
of the different multiple quantum coherences as a function of the
coherence order $\Delta M$ for different evolution times. Starting
from an initial cluster size $K_{0}=141$ for two different perturbations
strengths ($p=0.080$ in panel a, $p=0.025$ in panel b). In this
figure, the $A(\Delta M)$ values are not normalized. We find that
the width of the MQC distributions contracts as a function of time
in panel a), but expands in panel b). Like in the case where we start
from size $K_{0}=1$, this expansion does not continue indefinitely,
but it saturates. Similarly, the contraction also reaches an equilibrium
value. This is evident at the longest evolution times, where the gradients
of the MQC distributions, which give the cluster sizes, are parallel
in the semi-log scale representation. %

\begin{figure}
\includegraphics[bb=80bp 20bp 712bp 612bp,clip,scale=0.35]{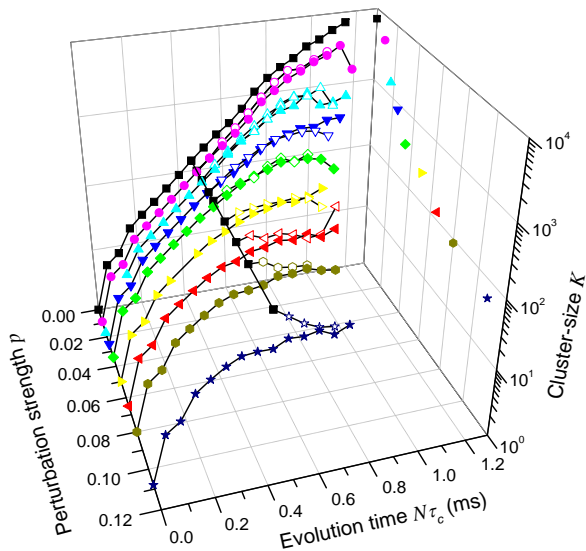}

\caption{(Color online) Time evolution of the  cluster-size starting from two
different initial states as a function of the perturbation strength.
Filled symbols are evolutions from an uncorrelated initial state ($K_{0}=1$)
and the empty symbols start from an initial state with $K_{0}=141$
correlated spins (marked with black squares joined with a solid line).
The equilibrium cluster sizes are represented by symbols in the $K-p$
plane.}

\label{Flo:cluster-size-k0fix-diffp} %
\end{figure}

To quantify this, we determined the width of the MQC spectra and from
that the cluster-size. Figure \ref{Flo:cluster-size-k0fix-diffp}
shows the evolution of the cluster-size when we start from the initial
value $K_{0}=141$ for different perturbation strengths (empty symbols).
The solid symbols represent the cluster-size evolution starting from
$K_{0}=1$, \emph{i.e.}, the curves shown in Fig. \ref{fig:Kvst}.
For all perturbation strengths, the time series for a given perturbation
strength converge to the same limiting value, independent of the initial
condition. Thus, we find that the evolution leads to a limiting cluster
size that varies with the perturbation strength $p$, but does not
depend on the initial condition $K_{0}$. The limiting cluster sizes
are represented by the symbols on the $K-p$ plane in Fig. \ref{Flo:cluster-size-k0fix-diffp}.

\begin{figure}
\includegraphics[bb=10bp 20bp 590bp 550bp,clip,scale=0.4]{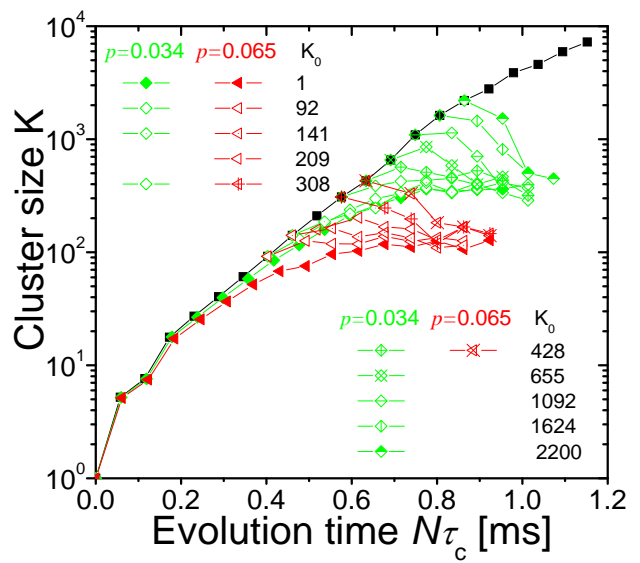}

\caption{(Color online) Time evolution of the  cluster size starting from different
initial sates. Filled symbols are evolutions from an uncorrelated
initial state for two different perturbation strengths given in the
legend. Empty symbols start from an initial state with $K_{0}$ correlated
spins.}

\label{fig:localization_convergence} %
\end{figure}

This is again confirmed by the results shown in Fig. \ref{fig:localization_convergence},
which summarizes the results for additional initial values and two
perturbation strengths, $p=0.034$ and $p=0.065$. The filled symbols
correspond to uncorrelated initial states ($K_{0}=1)$ and the empty
symbols to various initial cluster sizes. For a given perturbation
strength, the size of the spin clusters tends towards the same limiting
value, independent of the initial condition.

\section{Localization size vs. perturbation strength}

\subsection{Experimental evidence}

\begin{figure}
\includegraphics[bb=58bp 21bp 718bp 522bp,scale=0.3]{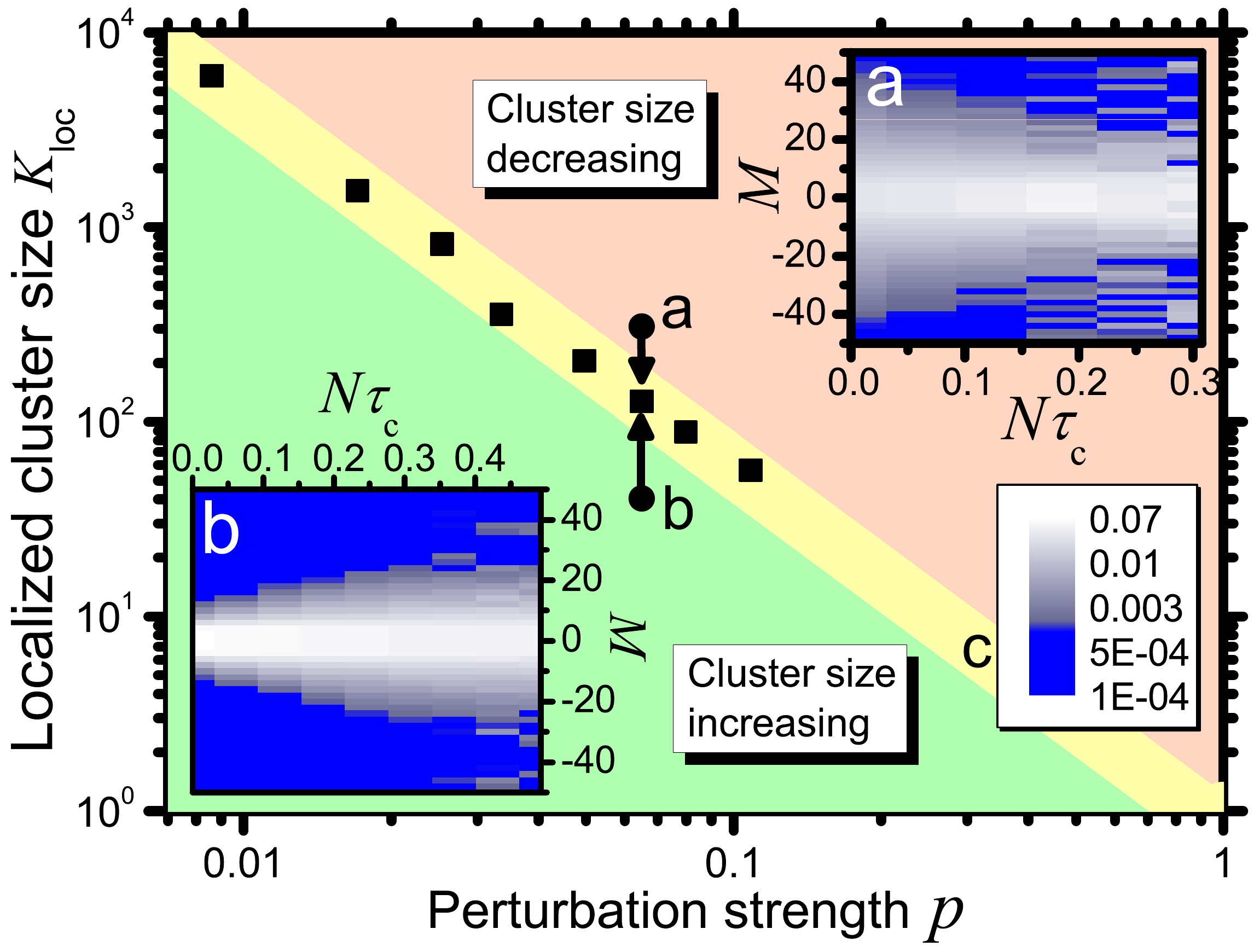}

\caption{(Color online) Localized cluster size $K_{\mathrm{loc}}$ (square
symbols) of correlated spins versus the perturbation strength $p$.
Three dynamical regimes for the evolution of the cluster size are
identified depending of the number of correlated spins compared with
the perturbation dependent localization value: a) a cluster size decreases,
b) a cluster size increases c) stationary regime.}

\label{fig:localizednumber} %
\end{figure}

According to Figs. \ref{fig:Kvst}, \ref{Flo:cluster-size-k0fix-diffp}
and \ref{fig:localization_convergence}, the size of the dynamical
equilibrium clusters decreases with increasing strength of the perturbation.
For a quantitative analysis of this dependence, we determined the
size of the localized clusters from the data shown in Figs. \ref{fig:Kvst},
\ref{Flo:cluster-size-k0fix-diffp} and \ref{fig:localization_convergence}
and plotted them against the perturbation strength in Fig. \ref{fig:localizednumber}.
The diagonal line in the figure represents a linear fit to the experimental
data (yellow/light line) that gives $K_{\mathrm{loc}}\sim p^{-1.86\pm0.05}$,
\emph{i.e.}, the cluster size decreases almost proportionally to the
square of the perturbation strength. The error of the fit is indicated
by the line-width. The limiting value for $p=1$, $K_{\mathrm{loc}}\approx1$,
is consistent with the expectation that the system becomes completely
localized if the perturbation strength is significantly larger than
the unperturbed Hamiltonian.

The figure also summarizes the evolution of the cluster size before
the static (localized) size is reached: If the initial size is larger
than the stationary value for the given perturbation strength, $K_{0}>K_{\mathrm{loc}}$,
the cluster shrinks {[}inset (a) in the figure, above the diagonal{]}.
If it is smaller, $K_{0}<K_{\mathrm{loc}}$, the size increases with
time {[}inset (b), below the diagonal{]}.

\subsection{Phenomenological model}

A theoretical model that describes this behavior would be highly desirable,
but is beyond the scope of this paper. Instead, we describe here a
simple phenomenological model that summarizes the observed behavior.

For this purpose, we write the amplitudes of the MQC spectra as

\begin{equation}
A(\Delta M,t)=Ae^{-\Delta M/\sqrt{K(t)}}.\label{eq:MQShape}\end{equation}
 The exponential form agrees with the experimental data (see, e.g.
Figs. \ref{Flo:Free_evolution} and \ref{Flo:MQC_spectrums_dyn_eq}).
According to Fig. \ref{fig:Growth}, the growth of the cluster size
is exponential during a large part of the experiment. It may then
be described by a differential equation of the type

\begin{equation}
\frac{dK}{dt}=\alpha K,\end{equation}
 where $\alpha$ is the growth rate, which is proportional to the
second moment of the NMR resonance line \cite{zobov_second_2006}.

To derive an equation of motion for the perturbed evolution, we first
consider the effect of the perturbation alone. As we have shown before
\cite{Suter04}, the interaction with the environment causes a decay
of the MQC amplitudes that depends on the MQC order $\Delta M$. It
was shown in Ref. \cite{Suter04} that over the range of our measurements
the decay rate is almost $\propto\Delta M$. The decay of the MQ amplitudes
during a time $\tau$ is thus $\delta A(\Delta M)\propto-A(\Delta M)\Delta Mpb\tau,$
where $p$ is the perturbation strength introduced above and $b$
is the decay rate of single quantum coherences under the full perturbation,
$p=1$.

During a short time $\tau$, the MQC spectrum evolves thus from (\ref{eq:MQShape})
to

\[
A(\Delta M,t+\tau)=Ae^{-\Delta M\left(1/\sqrt{K_{1}(t+\tau)}+pb\tau\right)}.\]
 Here, $K_{1}=K_{0}(1+\alpha\tau)$ is the size that the cluster size
would reach, starting from $K_{0}$ in the absence of the perturbation.
We can rewrite this as

\[
A(\Delta M,t+\tau)=Ae^{-\Delta M/\sqrt{K_{1}^{'}(t+\tau)}}\]
 with

\begin{align}
K_{1}^{'} & =\frac{K_{0}(1+\alpha\tau)}{\left(1+\sqrt{K_{0}(1+\alpha\tau)}pb\tau\right)^{2}}\nonumber \\
 & =\frac{K_{0}(1+\alpha\tau)}{1+2\sqrt{K_{0}(1+\alpha\tau)}pb\tau+K_{0}(1+\alpha\tau)(pb\tau)^{2}},\end{align}
 where for short $\tau$

\begin{align}
K_{1}^{'} & \approx\frac{K_{0}(1+\alpha\tau)}{1+2\sqrt{K_{0}(1+\alpha\tau)}pb\tau}\nonumber \\
 & \approx K_{0}(1+\alpha\tau)(1-2\sqrt{K_{0}(1+\alpha\tau)}pb\tau).\end{align}
 We now look for the stationary solution where $K_{1}'=K_{0}:$\begin{equation}
K_{1}^{'}=K_{0}=K_{0}(1+\alpha\tau)(1-2\sqrt{K_{0}(1+\alpha\tau)}pb\tau).\end{equation}
 If we consider only terms of $\mathcal{O}(\tau)$ in the infinitesimal
time $\tau$, this is equivalent to \begin{equation}
(\alpha-2\sqrt{K_{0}}pb)=0.\end{equation}
 Solving for $K_{0}$, we find the stationary cluster size as \begin{equation}
K_{\mathrm{loc}}=\left(\frac{\alpha}{2bp}\right)^{2}.\end{equation}
Considering the simplicity of this phenomenological model, this result
agrees reasonably well with the experimentally observed behavior.

\section{\textit{\emph{Discussion and Conclusions}}}

As a step towards improved understanding of the evolution of large
quantum systems, we have studied the spreading of information in a
system of nuclear spins. Starting with single qubits, the information
can spread to clusters of several thousand qubits. While decoherence
is well known to limit the time for which quantum information can
be used, we focused here on its effect on the distance over which
a quantum state can be transferred. For that purpose, a locally stored
state was left to evolve in a 3D spin-coupling network. Using standard
NMR techniques we generated a Hamiltonian $\mathcal{\widehat{H}}_{0}$
that spreads the information and allows one to quantify the size of
the resulting cluster of correlated spins. By comparing this unperturbed
evolution with the evolution where a perturbation Hamiltonian of variable
strength is added to $\mathcal{\widehat{H}}_{0}$ , we showed that
the information becomes localized on a distance scale that decreases
with increasing perturbation strength. 

Our experimental results demonstrate that a common dynamic equilibrium
size of the localized state is achieved independent of the initial
state consisting of different numbers of correlated spins. We developed
a phenomenological model to describe these effect, which may be attributed
to the competition between the spreading evolution driven by $\mathcal{\widehat{H}}_{0}$
and its systematic reduction due to decoherence induced by the perturbation.
These results are related with theoretical predictions on similar
systems that indicate a slow down of the spreading \cite{Burrell2007}
and can induce localized states with a finite cluster-size \cite{ZnidariC2008}.
Further extensions of these previous works could show if a dynamical
equilibrium independent of the initial state appears as in our experiments.
These previous works have considered disorder as the perturbation
that produces Anderson localization. 

The results presented here showed a transition in the spin dynamics
from an indefinite spreading to a localized dynamics. A spin far away
from the spin where the initial condition is stored would receive
excitation at some time if the perturbation is below a critical value,
however, if the perturbation exceeds this threshold no excitation
will arrive at this site. Our experiments show  a transition for
the cluster-size dynamics: if the cluster-size of the initial state
exceeds the localization value, the it shrinks until it reaches the
equilibrium value, but when the initial cluster size is lower than
the localization value, it grows.

These results may also be connected to our earlier findings that the
decoherence rate of quantum states with many correlated qubits increases
with the size of the system \cite{Suter04} as $\sqrt{K}$, indicating
that larger systems are more sensitive to perturbations. This increasing
of the decoherence rate, as the system size increases, balances the
tendency of the system to spread. 
\begin{acknowledgments}
This work was supported by the DFG through Su 192/24-1. GAA thanks financial support from an Alexander von Humboldt Research
Fellowship. We thank Marko Lovric, Hans Georg Krojanski and Ingo Niemeyer
for helpful discussions and technical support. 
\end{acknowledgments}
\bibliographystyle{apsrev} \bibliographystyle{apsrev} \bibliographystyle{apsrev}
\bibliography{bibliography,localization_long}

\end{document}